\documentclass[global,twocolumn]{revtex4-1}
\usepackage{amsmath,amssymb}
\usepackage{graphicx}
\usepackage{color}
\begin{document}
\title{Entangling quantum gate in trapped ions via Rydberg blockade}
\author{Weibin Li$^{1,2}$}
\author{Igor Lesanovsky$^1$}
\address{$^1$School of Physics and Astronomy, The University of Nottingham, Nottingham, NG7 2RD, United Kingdom}
\address{$^2$School of Physics, Huazhong University of Science and Technology, Wuhan 430074, China}
\date{\today}
\begin{abstract}
We present a theoretical analysis of the implementation of an entangling quantum gate between two trapped Ca$^+$ ions which is based on the dipolar interaction among ionic Rydberg states. In trapped ions the Rydberg excitation dynamics is usually strongly affected by mechanical forces due to the strong couplings between electronic and vibrational degrees of freedom in inhomogeneous electric fields. We demonstrate that this harmful effect can be overcome by using dressed states that emerge from the microwave coupling of nearby Rydberg states. At the same time these dressed states exhibit long range dipolar interactions which we use to implement a controlled adiabatic phase gate. Our study highlights a route towards a trapped ion quantum processor in which quantum gates are realized independently of the vibrational modes.
\end{abstract}
\maketitle

\section{Introduction}
A central effort in trapped ion quantum computation is the realization of controllable interactions between spatially separated qubits~\cite{divincenzo_2000}. This is key to establishing qubit entanglement for quantum information processing~\cite{schmidt-kaler_realization_2003,leibfried_experimental_2003,benhelm_towards_2008} and for transferring quantum information between remote qubits~\cite{brown_coupled_2011,harlander_2011}. In trapped ion quantum computation an effective qubit-qubit interaction~\cite{cirac_quantum_1995,sorensen_quantum_1999,milburn_ion_2000} is typically engineered by a state-dependent laser coupling of ions to quantum harmonic oscillations (phonons) of the ion crystal~\cite{porras_effective_2004,lee_phase_2005}. This use of phonons as quantum bus is currently limited to rather small arrangements of ions as the vibrational mode structure of a crystal becomes increasingly dense and complex as the number of ions grows. Executing gates in large ion crystals therefore can become very slow if one requires individual phonon modes to be spectroscopically resolved as proposed in many gate schemes~\cite{cirac_quantum_1995,sorensen_quantum_1999,milburn_ion_2000}. Hence, when scaling up a trapped ion quantum computer, the resulting slow gate speeds make it challenging to maintain quantum coherence within qubit which typically have a finite coherence time.

An alternative to the phonon induced effective interaction among qubits are long range dipolar interactions which are currently much studied in the context of neutral atoms \cite{saffman_quantum_2010}. These interactions occur when two atoms are excited to electronically high-lying (Rydberg) states~\cite{gallagher_rydberg_1988}. Depending on the specific Rydberg state the corresponding interaction strength can reach tens of MHz over a distance of several micrometers. This has pronounced consequences for the excitation dynamics of atoms which are excited from electronically low-lying (ELL) states to Rydberg states. In the extreme case it can lead to the so-called dipole blockade effect \cite{Lukin:2001}, i.e. the suppression of multiple excitation of Rydberg atoms within a certain volume~\cite{comparat_2010,saffman_quantum_2010,Dudin_Rabi_2012}. This phenomenon has been extensively explored in the context of many-body physics and it has been shown to be a central ingredient for the realization of two-qubit gates~\cite{jaksch_fast_2000}. Recently, the experimental demonstration of two-atom entanglement~\cite{wilk:2010} and a controlled-NOT quantum gate~\cite{Isenhower:2010} using the dipole blockade between Rydberg atoms has been reported.

Motivated by this, attempting to integrate dipolar interaction into trapped ion systems has generated considerable interest~\cite{muller_trapped_2008}. In terms of electronic structure, an atom and a singly charged ion in a Rydberg state share many similarities. One would therefore intuitively think that one could simply migrate the ideas developed for neutral atoms to the trapped ions. However, dramatic differences between neutral Rydberg atoms and trapped Rydberg ions emerge in both the motional and electronic dynamics. For neutral atoms, one commonly encounters situations where the atomic motion is frozen during the course of the Rydberg excitation~\cite{anderson_resonant_1998,mourachko_many-body_1998}. This condition can typically not be fulfilled for trapped ions and coherent couplings between electronic states and phonon motions are vitally important for understanding the trapped ion excitation dynamics.

\begin{figure}[h]
\centering
\includegraphics[width=3.2in]{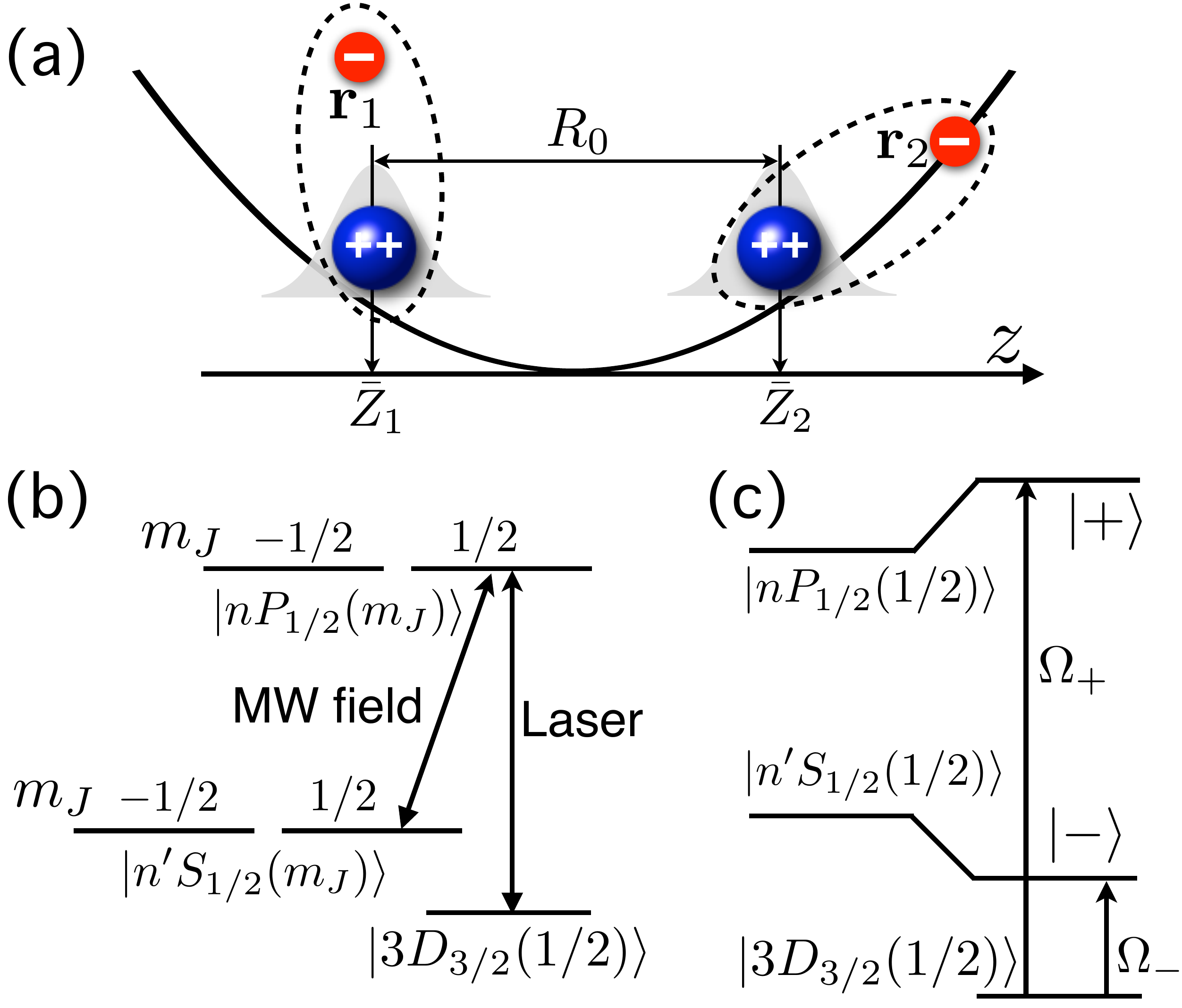}
\caption{(a) System schematics. Two Ca$^+$ ions form a crystal along the $z$-axis of the linear Paul trap. The equilibrium position of the $j$-th ion is $\{0,0,\bar{Z}_j\}$. In Rydberg states the size of the valance electron orbit becomes comparable or even larger than the width of the ionic wave packet (sketched as Gaussian in gray). The two ions (separated by $R_0$) interact through a dipolar interactions once excited to Rydberg states. (b) Relevant electronic levels used to describe the laser excitation and the MW dressing. A laser couples the low-lying $D$-state with the target Rydberg state, $|nP_{1/2}(1/2)\rangle$. The MW field couples the $|nP_{1/2}(1/2)\rangle$-state with a Rydberg $|n'S_{1/2}(1/2)\rangle$-state. (c) For a strong MW field, the dressed Rydberg states are separated by a large energy splitting given by the microwave Rabi frequency. Here, the selective laser addressing of individual MW dressed Rydberg state is possible.}
\label{fig:level}
\end{figure}
In Rydberg states the weakly bound electron can strongly couple to the ionic vibration. This is rooted in the fact that the Rydberg ion can not be regarded as a point-like particle but rather a composite object~\cite{muller_trapped_2008}: The orbital length of the Rydberg electron scales as $n^2$~\cite{gallagher_rydberg_1988} (with $n$ the principal quantum number of the Rydberg state). This length can be tens of nm at large $n$, which is several times larger than the typical oscillator length corresponding to the ionic vibration, $\sim 10$ nm (the situation is depicted in Fig.~\ref{fig:level}a). The strong coupling not only affects the electronic dynamics but also generates an additional ponderomotive potential for the external ionic motion~\cite{Li_crystal_2012}. This potential is proportional to Rydberg polarizability ($\sim n^7$) and thus modifies significantly the Rydberg ion trapping potential compared to ions in ELL states. This affects the laser excitation of ions from ELL states to a Rydberg state through the emergence of non-trivial Franck-Condon (FC) factors which characterize the overlap of phonon modes belonging to the two different potential surfaces. In general this leads to the mixing of vibrational state during the laser excitation which can cause qubit decoherence.

In addition a second issue arises, namely that the dipolar interaction between Rydberg ions is generally weaker than between neutral Rydberg atoms. This can be seen by the fact that the van der Waals (vdW) interaction (which typically is used to establish interactions among Rydberg atoms) scales as $1/\mathcal{Z}^6$ where $\mathcal{Z}$ is the net charge of the ionic core. This scaling immediately shows that the vdW interaction strength of Rydberg ions (with $\mathcal{Z}=+2$) is relatively weak compared to neutral atoms (with $\mathcal{Z}=+1$). Indeed, we will later show for the case of Ca$^+$ and typical trap parameters that it is not advantageous to use the vdW interaction for realizing two-qubit gates.

In this work we show how to overcome these two problems. Our solution relies on the application of a microwave (MW) field which couples ionic Rydberg states which have a polarizability of opposite signs (see level scheme depicted in Fig.~\ref{fig:level}b). The polarizability of the emerging dressed Rydberg states (Fig.~\ref{fig:level}c) can be dynamically switched off. In this case the difference of the potential surface of ions in the Rydberg state and the ELL states is lifted and the Rydberg excitation decouples from the phonon dynamics. In addition, the large permanent (rotating) dipole moment of the MW dressed Rydberg states generates a dipole-dipole (DD) interaction which is substantially stronger than the vdW interaction. This permits the implementation of a fast and robust quantum gate.

The paper is organized as follows. In Sec.~\ref{system}, we present and analyze the Hamiltonian governing the coupled vibrational and electronic dynamics of two Ca$^+$ ions trapped  in a linear Paul trap in the presence of laser and MW fields. Here we will elaborate in more detail on the aforementioned MW control of the Rydberg excitation and interaction strength. In  Sec.~\ref{gate}, we discuss the implementation of a controlled two-qubit phase gate relying on the DD interaction between MW dressed Rydberg states. We conclude in Sec.~\ref{conclusion}.

\section{System and Hamiltonian}
\label{system}
We consider two Ca$^+$ ions of mass $M$ trapped in a linear Paul trap whose electric potential is given by
$\Phi({\bf r},t)=\alpha\cos\Omega t(x^2-y^2)-\beta(x^2+y^2-2z^2)$. Here $\alpha$ and $\beta$ are the electric field gradients of the radio frequency and static field, respectively, and $\Omega$ is the oscillation frequency of the radio frequency field. At the trap center, this field configuration gives rise to an effective harmonic ponderomotive harmonic potential~\cite{leibfried_quantum_2003} for the ions with  axial and longitudinal trap frequencies $\omega_{\rho}=\sqrt{2[(e\alpha/M\Omega)^2-e\beta/M]}$ and $\omega_Z=2\sqrt{e\beta/M}$, respectively ($e$ the elementary charge). The Hamiltonian of ion $j\,(j=1,2)$ is~\cite{muller_trapped_2008}
\begin{equation}
H_j= H_{\rm{CM}}(\textbf{R}_j)+H_{\rm{e}}(\textbf{r}_j)+H_{\rm{ec}}(\textbf{R}_j,\textbf{r}_j)+V_{\rm{ext}},
\label{Hcen}
\end{equation}
where $H_{\rm{CM}}(\textbf{R}_j)$ describes the harmonic oscillation of the center-of-mass (CM) coordinate ($\mathbf{R}_j=\{X_j,Y_j,Z_j\}$) of the respective ion. Furthermore, $H_{\rm{e}}(\textbf{r}_j)$ is the Hamiltonian of the valence electron ($\mathbf{r}_j=\{x_j,y_j,z_j\}$) and $H_{\rm{ec}}(\textbf{R}_j,\textbf{r}_j)$ is the electron-CM coupling Hamiltonian of the $j$-th ion, respectively. The explicit form of the individual terms is given by
\begin{eqnarray}
\label{eq:cmharmonic}
H_{\rm{CM}}(\textbf{R}_j)&=& \frac{\mathbf{P}_j^2}{2M} +\frac{M}{2}\left[\omega_{\rho}^2(X_j^2+Y_j^2)+\omega_Z^2 Z_j^2\right], \\
\label{eq:electron}
H_{\rm{e}}(\textbf{r}_j)&=&\sum_{\mathbf{L}}\epsilon_{\mathbf{L}}|\mathbf{L}\rangle\langle \mathbf{L}|+H_{\rm{et}}(\textbf{r}_j), \\
\label{eq:electroncm}
H_{\rm{ec}}(\textbf{R}_j,\textbf{r}_j)&=& -2e\left[\alpha\cos\omega t(X_jx_j-Y_jy_j)\right.\\
&-&\left.\beta(X_jx_j+Y_jy_j-2Z_jz_j)\right],
\end{eqnarray}
where $H_{\rm{et}}(\textbf{r}_j)=-e\Phi(\textbf{r}_j,t)$ is the coupling between the valence electron and the electric field of the Paul trap. To label the electronic states we introduce the multi-index $\mathbf{L}=\{n,\,L,\,J,\,m_J\}$, where $n,\, L,\, J$ are the principle, angular, total angular quantum number and  $m_J$ is the projection of $J$ on the quantization axis ~\cite{schmidt-kaler_rydberg_2011} and $\epsilon_{\mathbf{L}}$ is the respective state energy. Finally, the term $V_{\rm{ext}}$ describes the interaction of the ion with external laser and microwave fields, whose form will be given later. Note that the effect of micromotion~\cite{leibfried_quantum_2003} has been neglected in this description.

The constituents of the two ions interact with the Coulomb interaction. In linear ion traps, the typical inter-ion separations are about $5\,\mu$m, which is far larger than the characteristic length of both electron orbits and ionic vibration. This allows us to Taylor expand the Coulomb interaction in terms of the inter-ion separation $R_{0}$~\cite{muller_trapped_2008}
\begin{eqnarray}
V(\mathbf{R}_1,\mathbf{R}_2,&&\mathbf{r}_1,\mathbf{r}_2)/C_0  \approx  \frac{1}{R_0} + \frac{\mathbf{n}_{12}\cdot (\mathbf{r}_1 -\mathbf{r}_2)}{R_0^2}\nonumber\\
&&+ \frac{r_1^2 -3(\mathbf{n}_{12}\cdot \mathbf{r}_1)^2 + r_2^2 -3(\mathbf{n}_{12}\cdot \mathbf{r}_2)^2}{2 R_0^3} \nonumber \\
& & + \frac{\mathbf{r}_1\cdot \mathbf{r}_2-3 (\mathbf{n}_{12}\cdot \mathbf{r}_1) (\mathbf{n}_{12}\cdot \mathbf{r}_2)}{R_0^3}\cdots
\label{eq:multipolar}
\end{eqnarray}
where $C_0= e^2/4\pi\epsilon_0$ with $\epsilon_0$ being the vacuum permittivity. We have furthermore used $R_0=|\bar{\mathbf{R}}_1-\bar{\mathbf{R}}_2|$ and $\mathbf{n}_{12}=(\bar{\mathbf{R}}_1-\bar{\mathbf{R}}_2)/R_0$ with $\bar{\mathbf{R}}_j$ being the equilibrium position of the $j$-th ion. On the right hand side of Eq.~(\ref{eq:multipolar}), the first term is the Coulomb interaction between the two singly charged ions. Higher order terms in the expansion give contributions due to electron-charge and electron-electron interaction. The second and third term are the dipole-charge and quadrupole-charge interaction. The fourth term is the dipole-dipole interaction.

These higher order terms have different impacts on the electronic and ionic dynamics. For the linear crystal, the dipole-charge interaction cancels the $z$-component of the electron-CM coupling. Hence the equilibrium positions of the ions are unaffected by a change of the electronic state. The quadrupole-charge interaction modifies the electronic Hamiltonian as $H_e'(\textbf{r}_j)=H_e(\textbf{r}_j)+C_0(x_j^2+y_j^2-2z_j^2)/(2R_0^3)$~\cite{muller_trapped_2008}. Consequently the electronic energies are shifted according to $\epsilon_{\mathbf{L}}'=\epsilon_{\mathbf{L}}+\delta'_{\rm{e}}$ where the corresponding energy shift $\delta'_{\rm{e}}$ can be calculated via second order perturbation theory as discussed in Ref. ~\cite{schmidt-kaler_rydberg_2011}.

As highlighted in the introduction, the importance of certain terms in the two-ion Hamiltonian strongly depends on the considered electronic states. For example, in ELL states, the electron-CM and electron-trap coupling as well as higher order terms in Eq.~(\ref{eq:multipolar}) can be safely neglected. However, due to the strong scaling of characteristic quantities such as the polarizability as a function of the principle quantum number $n$~\cite{gallagher_rydberg_1988}, terms whose effect is negligible in case of ELL states become important for the system dynamics when ions are excited to Rydberg states. This will have an impact, e.g., on the laser excitation dynamics. In the following we will discuss this in detail.

\subsection{Hamiltonian of ions in ELL states}
In an ELL state, the dynamics of the valence electron is hardly affected by the electric fields of the trap and the electronic and phonon dynamics are essentially decoupled. At sufficiently low temperature the ions form a Wigner crystal as a result of interplay between the Coulomb repulsion and the trap confinement~\cite{james_quantum_1998}. For our linear two-ion crystal the equilibrium positions are
\begin{eqnarray}
\label{eq:xyposition}
\bar{X}_j&=&\bar{Y}_j=0 \\
\label{eq:zposition}
-\bar{Z}_1&=&\bar{Z}_2=\left(\frac{C_0}{16e\beta}\right)^{1/3}.
\end{eqnarray}
When displaced from the equilibrium positions the two ions couple with each other through the Coulomb interaction. The resulting coupled vibrations are described in terms phonon modes with the Hamiltonian
\begin{equation}
H_{\mathrm{v}}=\sum_{\chi=X,Y,Z}\sum_{j=1,2}\hbar\omega_{\chi,j}a^{\dagger}_{\chi,j}a_{\chi,j}.
\end{equation}
Here $a^{\dagger}_{\chi,j}$ ($a_{\chi,j}$) is the creation (annihilation) operator of the $j$-th phonon mode along the $\chi$-axis. The phonon frequencies $\omega_{\chi,j}$ are calculated by diagonalizing the Hessian matrix, $\sum_m\mathcal{H}_{mn}^{(g,\chi)}\mathbf{A}_m^{(\chi,j)}=\omega_{\chi,j}^2\mathbf{A}_n^{(\chi,j)}$ ($\mathbf{A}^{(\chi,j)}$ denotes the eigenvector of the respective phonon mode) with the matrix elements
\begin{equation}
\mathcal{H}_{mn}^{(g,\chi)}=\left\{\begin{array}{ll}\omega_{\chi}^2-\frac{c_{\chi}}{(2\bar{Z}_1)^3}, & n=m \\ \frac{c_{\chi}}{(2\bar{Z}_1)^3},&n\neq m \end{array}\right.
\nonumber
\end{equation}
where $c_{X}=c_{Y}=1$, $c_{Z}=-2$.

\subsection{Dynamics of the laser excitation of Rydberg states}
\label{laserexcitation}
Let us now study the dynamics of the laser excitation of ions from the ELL state to Rydberg states. Specifically we consider that the Ca$^+$ ions are excited from the low-lying $|D\rangle=|3D_{3/2}(1/2)\rangle$ to the Rydberg $|P\rangle=|nP_{1/2}(1/2)\rangle$-state (see Fig. \ref{fig:level}b) via a single photon transition as it can be achieved with a vacuum ultraviolet laser (a thorough discussion of such vacuum ultraviolet laser in the context of ionic Rydberg excitations can be found in Refs.~\cite{schmidt-kaler_rydberg_2011,kolbe_2012}).

To understand the excitation process it is instructive to characterize first the effective trapping potential experienced by a Rydberg ion. In the Rydberg state, the large electron-CM coupling gives rise to an additional ponderomotive potential to the CM motion. The details of the derivation can be found in Refs.~\cite{Li_crystal_2012,li_mode_2013}. Note furthermore, that similar state-dependent modifications of the trapping potential occur also in case of neutral Rydberg atoms in magnetic traps~\cite{lesanovsky_magnetic_2005}. In the $|P\rangle$-state, the additional trapping potential experienced by the $j$-th ion is
\begin{equation}
V_a(\textbf{R}_j)\approx-e^2\alpha^2\mathcal{P}_{P}(X_j^2+Y_j^2),
\label{eq:addpotential}
\end{equation}
where $\mathcal{P}_P$ is the polarizability in the Rydberg state. Here terms containing the static gradient $\beta$ have been neglected with respect to those containing the gradient $\alpha$ of the radio-frequency field as typically $\alpha\gg\beta$ in linear ion traps. Since $\mathcal{P}_{P}\propto n^7$, the additional potential can result in a major modification of the harmonic confinement which strongly affects the phonon mode structure. This can be exploited for the dynamical mode shaping within ion chains~\cite{li_mode_2013}. The Hessian matrix characterizing the transverse phonon modes in the Rydberg state is given by
\begin{eqnarray}
\mathcal{H}_{mn}^{(\mathcal{P},\chi)}=\left\{\begin{array}{ll}\omega_{\chi}^2-2e^2\alpha^2\mathcal{P}_{P}-\frac{1}{(2\bar{Z}_1)^3}, & n=m \\ \frac{1}{(2\bar{Z}_1)^3},&n\neq m \end{array}\right.
\end{eqnarray}
with $\chi=X,\,Y$. The eigenvector $\mathbf{B}^{(\mathcal{P},\chi,j)}$ and eigenfrequency $\nu_{\mathcal{P},\chi,j}$ of the phonon modes are obtained by solving $\sum_m\mathcal{H}_{mn}^{(\mathcal{P},\chi)}\mathbf{B}_m^{(\mathcal{P},\chi,j)}=\nu_{\mathcal{P},\chi,j}^2\mathbf{B}_n^{(\mathcal{P},\chi,j)}$. The respective (electronic state-dependent) phonon Hamiltonian is
\begin{equation}
H_{\mathrm{v}}^{(\mathcal{P})}=\sum_{\chi=X,Y}\sum_{j=1,2}\hbar\nu_{\mathcal{P},\chi,j}b^{\dagger}_{\mathcal{P},\chi,j}b_{\mathcal{P},\chi,j},
\end{equation}
with $b^{\dagger}_{\mathcal{P},\chi,j}$ and $b_{\mathcal{P},\chi,j}$ being the phonon creation and annihilation operators, respectively.
\begin{figure}[h]
\centering
\includegraphics[width=3.2in]{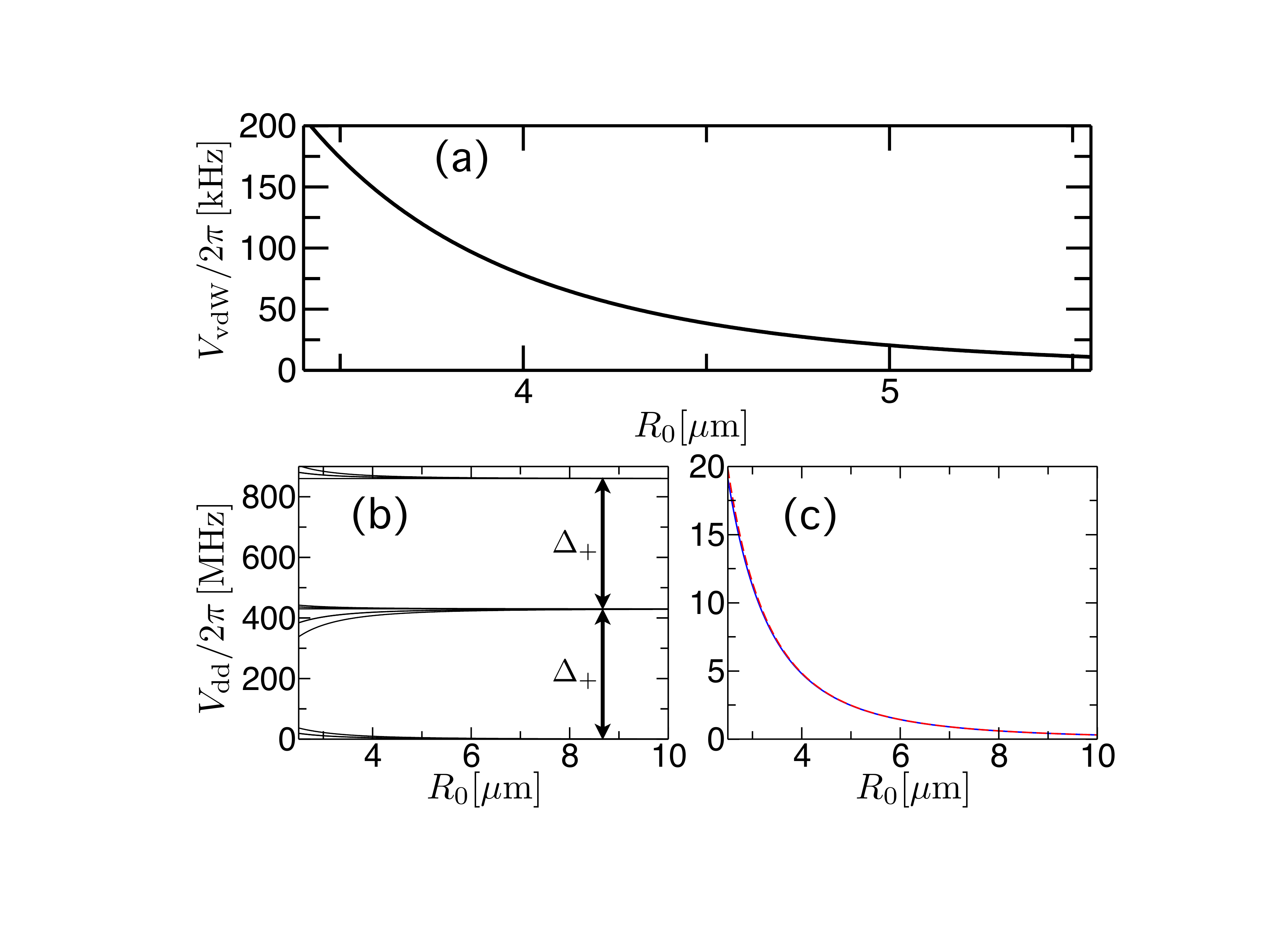}
\caption{(a) Van der Waals interaction between two ions in the Rydberg $|65P_{1/2}(1/2)\rangle$ state as a function of the ion separation $R_0$. The total angular momentum projection of the two ion is 1, i.e. $m^{(1)}_J+m^{(2)}_J=1$. The vdW interactions for the ions in other Zeeman states (different projection quantum number $m^{(1)}_J+m^{(2)}_J$ but same $n$ and $J$) are virtually indistinguishable on the scale displayed in the figure. (b) DD interaction (as function of $R_0$) between ions in MW dressed Rydberg states. The MW field preserves the magnetic quantum number in the $|65P_{1/2}(m_{1/2})\rangle$-$|65S_{1/2}(m_{1/2})\rangle$ transition ($m_J=\pm1/2$). The dressed state (see the main text) are well defined as long as the mixing of Rydberg states of different $m_J$ by the DD interaction is negligible, which is guaranteed by the strong MW driving. The data displayed in the panel are in this strong MW driving regime. The parameters used are $\Omega_{\rm{MW}}=2\pi\times 400$ MHz, $\Delta_S=2\pi\times 136.074$ MHz and $\Delta_P=2\pi\times 293.957$ MHz. These parameters also result in a vanishing polarizability of the dressed Rydberg $|-\rangle$-state. (c) DD interaction as a function of $R_0$ in the electronic pair state $|--\rangle$. The solid curve is the full calculation and the dashed one displays the approximate result given by Eq. (\ref{eq:atlongrange}). Both curves are undistinguishable on the scale used in the figure. See text for details.}
\label{fig:longrangeinteraction}
\end{figure}

We are now in a position to investigate the laser excitation dynamics and how it is affected by the electronic state-dependent phonon modes. We assume that the excitation laser propagates along the $z$-axis and that it is polarized along the $y$-axis. The ion-laser interaction is described by the coupling Hamiltonian
\begin{equation}
V_L(t)=-eE_0(Z_j)\,[\mathbf{r}_i\cdot\hat{\varepsilon}_0]\cos\omega_0 t,
\label{eq:ionlaser}
\end{equation}
where $E_0(Z_j)$ is the strength of the laser field at the position of the $j$-th ion and $\omega_0$ is the laser frequency. In order to obtain an explicit expression for the laser coupling induced between the states $|D\rangle$ and $|P\rangle$ we transform into a rotating frame with the unitary transformation $U_{L}=P_D+e^{i\omega_0 t}P_P$, using the projection operators $P_D=|D\rangle\langle D|$, $P_P=|P\rangle\langle P|$. In the interaction picture and within the rotating wave and resonance approximation we arrive at the ion-laser interaction Hamiltonian~\cite{Li_crystal_2012}
\begin{equation}
\label{eq:laserhamiltonian}
H_{L} \approx \sum_{\substack{\chi=X,Y \\ [j],[k]}}\frac{\Omega(Z_j)}{2}K_{[j]}^{[k]}b^{\dagger}_{\mathcal{P},\chi,[k]}a_{\chi,[j]}\otimes |P\rangle\langle D|+\rm{H.c.},
\end{equation}
where $\Omega(Z_j)=E_0(Z_j)d_0$ is the laser Rabi frequency with $d_0=-e\langle P|y_j|D\rangle$ being the transition dipole moment between the $|D\rangle$- and $|P\rangle$-state. The coefficients $K_{[j]}^{[k]}$ are the FC factors and given by the overlap integrals between the vibrational modes in the potential surface corresponding to the $|D\rangle$- and $|P\rangle$-state. We furthermore have defined $b^{\dagger}_{\mathcal{P},\chi,[k]}=(b^{\dagger}_{\mathcal{P},\chi,1})^{k_1}(b^{\dagger}_{\mathcal{P},\chi,2})^{k_2}/\sqrt{k_1!k_2!}$, where $k_i$ is the phonon number of the $i$-th mode along the $\chi$ axis. A similar definition is used for the operators $a_{\chi,[j]}$ in Eq.~(\ref{eq:laserhamiltonian}).

From Eq.~(\ref{eq:laserhamiltonian}) it is evident that the Rydberg excitation depends strongly on the FC factors~\cite{sharp_franckcondon_1964} when the trapping potential is state-dependent. In general this results in a strong and rather intricate coupling between vibrational and electronic degrees of freedom. For the implementation of quantum gates this is not desirable and indeed for many gate schemes it is advantageous to have a trapping potential that is essentially independent of the electronic state~\cite{garcia_2003,duan_2004,garcia_coherent_2005}.

\subsection{MW dressing of Rydberg states}
\label{mwdressing}
In order to achieve such a trapping potential which does not depend on the electronic state, and therefore gives rise to trivial FC factors, i.e. $K_{[j]}^{[k]}=\delta_{jk}$, we do not work with bare Rydberg states. Instead we create dressed states by using a strong MW field which couples the Rydberg $|P\rangle$-state with a Rydberg $|S\rangle=|n',S_{1/2}(1/2)\rangle$ as shown in Fig.~\ref{fig:level}b. The states are chosen such that the signs of their repsective polarizabilities are opposite, i.e. $\mathcal{P}_{n'S}>0$ and $\mathcal{P}_{nP}<0$. With an appropriate choice of the MW coupling this permits the creation of dressed states with vanishing polarizability and thus removes the additional potential (\ref{eq:addpotential}) in the Rydberg state.

Let us now discuss the practical implementation of this idea. The interaction Hamiltonian of the $j$-th ion with the MW field is given by $V_{\rm{MW}}(\textbf{r}_j)=-eE_1\,[\textbf{r}_j\cdot \hat{\varepsilon}]\cos\omega_1 t$, with $E_1$, $\omega_1$ and $\hat{\varepsilon}$ the electric field strength, frequency and polarization (along $y$-axis). We use the unitary transformation $U_{t}=P_D+P_Pe^{i\omega_0 t}+P_Se^{(i\omega_0\pm\omega_1)t}$ to move into a rotating frame with respect to both the MW and the laser with $P_S=|S\rangle\langle S|$. In the unitary transformation, the $+$ corresponds to the situation $\epsilon_P<\epsilon_S$ and $-$ to the situation $\epsilon_P>\epsilon_S$. For concreteness we will assume $\epsilon_P>\epsilon_S$ in the following.

We consider the regime of a strong MW field~\cite{poyatos_1996}, i.e. the timescale related to the MW coupling is much shorter than that of the ionic motion and the ion-laser interaction. In this regime, the interaction of the MW with the $j$-th ion is described by~\cite{li_mode_2013}
\begin{eqnarray}
\label{eq:ionmw}
H_{\rm{MW}}(\textbf{r}_j)&=&\Delta_S|S \rangle_j \langle S|+\Delta_P|P\rangle_j \langle P|\nonumber\\
& + &\frac{\Omega_{\rm{MW}}}{2}(|S\rangle_j\langle P|+\rm{H.c.}),
\end{eqnarray}
where $\Delta_S=\epsilon_S'-(\omega_0-\omega_1)$, $\Delta_P=\epsilon_P'-\omega_0$, and $\Omega_{\rm{MW}}=E_1d_1$ is the MW Rabi frequency with $d_1=-e\langle P|y_j|S\rangle$ being the transition dipole moment between the Rydberg $|P\rangle$- and $|S\rangle$-state. Diagonalizing this Hamiltonian, we obtain two MW dressed Rydberg states
\begin{eqnarray}
|\pm\rangle_j&=&N_{\pm}\left(C_{\pm}|P\rangle_j+|S\rangle_j\right),
\end{eqnarray}
where $C_{\pm}=\frac{\Delta_-\pm\sqrt{\Omega_{\rm{MW}}^2+\Delta_-^2}}{\Omega_{\rm{MW}}}$ with $\Delta_{\pm}=\Delta_P\pm\Delta_S$ and $N_{\pm}=1/\sqrt{1+C_{\pm}^2}$ is the normalization constant. The dressed state energy is $E_{\pm}=\frac{\Delta_+}{2}\pm\frac{1}{2}\sqrt{\Omega_{\rm{MW}}^2+\Delta_-^2}$. The polarizability of the dressed state, $\mathcal{P}_{\pm}= N_{\pm}^2(C_{\pm}^2\mathcal{P}_{nP}+\mathcal{P}_{n'S})$, can be controlled by tuning the MW parameters. For example, for $n'=n$, $\mathcal{P}_{\pm}\approx 0$ when $|C_{\pm}|\approx 0.68$, i.e. the polarizability vanishes. When exciting such a dressed state with vanishing polarizability, the trapping potential of the Rydberg ion becomes identical with that of the ions in ELL states. Here, the FC factors~\cite{Li_crystal_2012} become trivial and the laser excitation is not different as compared to laser transitions driven among ELL states.

\subsection{Dipolar interaction between MW-dressed Rydberg ions}
\label{dipolarinteraction}

As discussed at the beginning of Sec.~\ref{system}, the operator describing the electron-electron interaction between two ions is given by
\begin{equation}
V_{\rm{dd}}(\textbf{R}_1,\textbf{R}_2)=C_0\frac{\mathbf{r}_i\cdot \mathbf{r}_j-3 (\mathbf{n}_{ij}\cdot \mathbf{r}_i) (\mathbf{n}_{ij}\cdot \mathbf{r}_j)}{R_0^3}.
\label{eq:ddpotential}
\end{equation}
For typical experimental parameters the resulting interaction energy is negligible for ions in ELL states. In Rydberg states this interaction can become significant and its actual functional form depends strongly on whether or not the electronic states possess a permanent dipole moment. We investigate these two situations in the following.

In the absence of MW dressing (see discussion in Sec.~\ref{laserexcitation}) the laser excites the Rydberg state $|P\rangle$ which possess no permanent dipole moment. As a result, the interaction energy shift between two ions excited to the  $|P\rangle$-state has the form of a van der Waals potential, $V_{\rm{vdW}}=C_6/R_0^6$, with $C_6$ being the dispersion coefficient (see Fig.~\ref{fig:longrangeinteraction}a). For $n=65$, we obtain $C_6\approx 2\pi\times0.3$ GHz $\mu m^6$, which results in an interaction shift of $2\pi\times 20$ kHz at $R_0=5\mu m$. For all practical purposes, e.g. the implementation of a two-qubit gate protocol, this interaction energy is too small.

This changes however in the presence of the MW field. The dressed Rydberg states of the ion exhibit a rotating dipole moment which leads to resonant exchange of MW photons between the two ions. This process gives rise to a DD interaction~\cite{comparat_2010} of the form
\begin{equation}
\label{eq:atlongrange}
V_{\rm{dd}}(\pm)\approx \frac{C_0}{R_0^3}(d_-^2\mathcal{P}_-+d_+^2\mathcal{P}_+),
\end{equation}
where $\mathcal{P}_{+}=|++\rangle\langle ++|$ (using the notation $|++\rangle=|+\rangle_1|+\rangle_2$) and $\mathcal{P}_{-}=|--\rangle\langle --|$ are the projection operators on the respective ion pair states. The interaction strength is determined by the parameter $d_{\pm}=N_{\pm}^2C_{\pm}|d_1|/e$. In order to derive expression (\ref{eq:atlongrange}) we have performed several approximations.  First, fast oscillating terms (with frequency $2\omega_1$) and DD couplings between the two dressed states are neglected. The latter is justified due to the large Autler-Townes splitting between the dressed state (illustrated in Fig.~\ref{fig:longrangeinteraction}b). Second, we neglected the $x$ and $z$ components in the DD interaction operator Eq.~(\ref{eq:ddpotential}) as these couplings are vanishingly small in the MW dressed state. To verify this second approximation, we numerically calculate the two-ion interaction potential by including both the MW driving [Eq.~(\ref{eq:ionmw})] and the DD interaction [(Eq.~(\ref{eq:ddpotential})]. As shown in Fig.~\ref{fig:longrangeinteraction}c, the potential in the pair $|--\rangle$ state based on this full calculation (without the aforementioned approximations) agrees well with that of the simplified calculation [Eq.~(\ref{eq:atlongrange})]. Hence we can reliably obtain the DD interaction strength from Eq.~(\ref{eq:atlongrange}). Using typical parameters, e.g. $\Omega_{\rm{MW}}=2\pi\times 400$ MHz, $\Delta_S=2\pi\times 136.074$ MHz and $\Delta_P=2\pi\times 293.957$ MHz, the DD interaction strength evaluated to $C_{3}(-)=C_0 d_-^2 \approx 2\pi \times 0.309$ GHz $\mu m^3$ for $n=n'=65$. For an ion separation of $R_0=5\mu$m, the DD interaction energy is $\approx 2\pi\times 2.5$ MHz, which is significantly larger than the Rydberg excitation Rabi frequency. In this parameter regime the simultaneous excitation of ions into Rydberg pair states is strongly suppressed. We will use this so-called dipole blockade in the following section for the implementation of a two-qubit entangling gate - the controlled phase gate.

\section{Implementation of a two-qubit phase gate}
\label{gate}
\begin{figure}[h]
\centering
\includegraphics[width=3.0in]{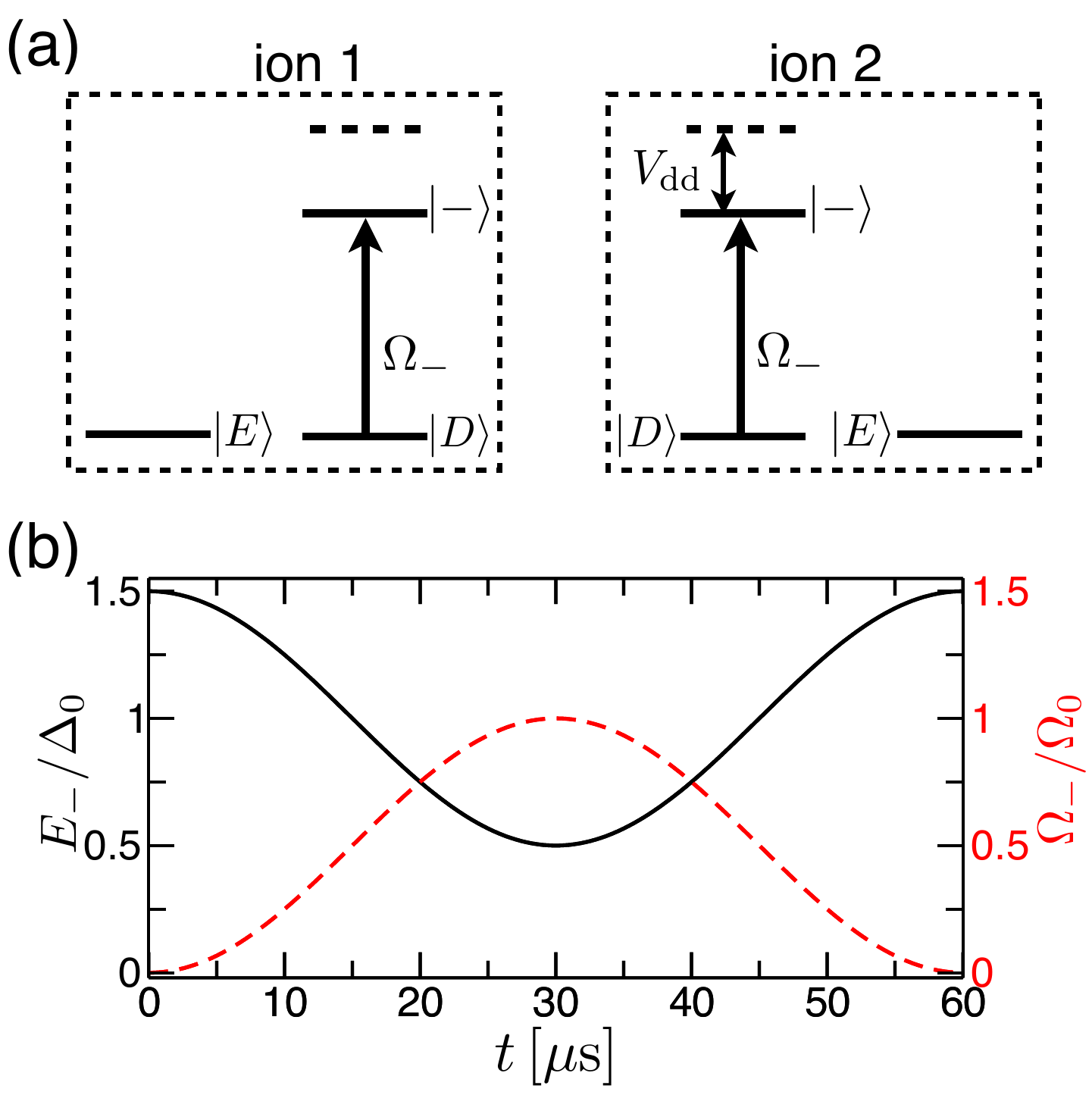}
\caption{(a) Level scheme used for implementing the two-qubit phase gate. $|D\rangle$ and $|E\rangle$ form the logical states of the qubit and the states $|D\rangle$ and $|-\rangle$ are coupled by a laser field with time-dependent detuning and Rabi frequency profile. (b) Laser pulse shape. Temporal variation of the Rabi frequency (dashed line) and detuning (solid line). The parameters are $\Omega_0=2\pi\times0.5$ MHz, $\Delta_0=2\pi\times 0.639$ MHz and $\tau=60\,\mu$s.}
\label{fig:gate_laser}
\end{figure}

Among the many existing protocols~\cite{saffman_quantum_2010}, we focus on an adiabatic scheme for the implementation of the phase gate. This scheme benefits from the fact that phonon excitation is largely negligible and that the laser addressing of individual ions is not required~\cite{jaksch_fast_2000}. The logical qubit states of each ion are the states $|D\rangle$ and a second ELL state $|E\rangle$ (e.g. the ground state $|4S\rangle$ of Ca$^+$). In order to implement the gate we use a Rydberg laser driving the $|D\rangle \leftrightarrow |-\rangle$ transition of each ion (the corresponding level in this gate scheme is depicted in Fig.~\ref{fig:gate_laser}a). We assume an excitation laser propagating along the trap axis, whose Rabi frequency is both time- and space-dependent, i.e. $\Omega(Z_j)=E_0(t)d_0\exp(ik_{\rm{L}}Z_j)$ where $k_{\rm{L}}$ is the wave number of the Rydberg laser. Such time dependence of the laser electric field can be achieved by varying the laser intensity. The effective Rabi frequency for the $|D\rangle\leftrightarrow|-\rangle$ transition is $\Omega_{-}(Z_j)=\Omega_{\rm{MW}}\Omega(Z_j)/\sqrt{4N_{-}^2(\Omega_{\rm{MW}}^2 +\Delta_-^2)}$ which can be further parameterized as $\Omega_{-}(Z_j)=\Omega_-(t)\exp(ik_{\rm{L}}Z_j)$ with $\Omega_-(t)=\Omega_{\rm{MW}}d_0E_0(t)/\sqrt{4N_{-}^2(\Omega_{\rm{MW}}^2 +\Delta_-^2)}$. In what follows we assume that only the axial CM phonon mode is coupled with the electronic dynamics through the spatial dependence of $\Omega(Z_j)$. We expand $\Omega_-(Z_j)$ in terms of the Lamb-Dicke parameter $\eta$ and truncate the expansion up to the first order of $\eta$, i.e.
$\Omega_-(Z_j)\approx\Omega_{-}(t)[1+i\eta(a_{z}^{\dagger}+a_{z})]$.
Here $a_z^{\dagger}$ ($a_z$) is the phonon creation (annihilation) operator of the axial CM mode, $\eta=k_{\rm{L}}\xi_{z}/\sqrt{2}$ is the Lamb-Dicke parameter and $\xi_{z}$ is the oscillator length of the CM mode. The resulting two-ion Hamiltonian is
\begin{eqnarray}
\label{eq:gatehamiltonian}
&&H\approx H_{\rm{v,z}}+\frac{C_3(-)}{R_0^3} \mathcal{P}_-+\\ \nonumber
&&\sum_{j=1,2}\left\{E_-(t)|-\rangle_j\langle-|+\frac{\Omega_{-}(t)}{2}[1+i\eta(a_{z}^{\dagger}+a_{z})]\sigma^{(j)}_++\rm{H.c.}\right\},
\end{eqnarray}
where $H_{\rm{v,z}}=\omega_za_{z}^{\dagger}a_{z}$, $\sigma_+=|-\rangle\langle D|$, $\sigma_-=\sigma_+^{\dagger}$ and $E_-(t)$ is the detuning of the Rydberg excitation laser frequency with respect to the$|D\rangle \leftrightarrow |-\rangle$ transition.
To realize the adiabatic phase gate, we consider a laser pulse whose Rabi frequency and detuning are time-dependent (see also Fig.~\ref{fig:gate_laser}b):
\begin{eqnarray}
\Omega_-(t)&=&\Omega_0\sin^2\left(\frac{\pi}{\tau}t\right),\nonumber\\
E_-(t)&=&\Delta_0\left[\frac{1}{2}+\cos^2 \left(\frac{\pi}{\tau}t\right)\right]\nonumber.
\end{eqnarray}
Here $\Omega_0$, $\Delta_0$ and $\tau$ (the duration of the gate laser pulse) are constants.
\begin{figure}[h]
\centering
\includegraphics[width=2.8in]{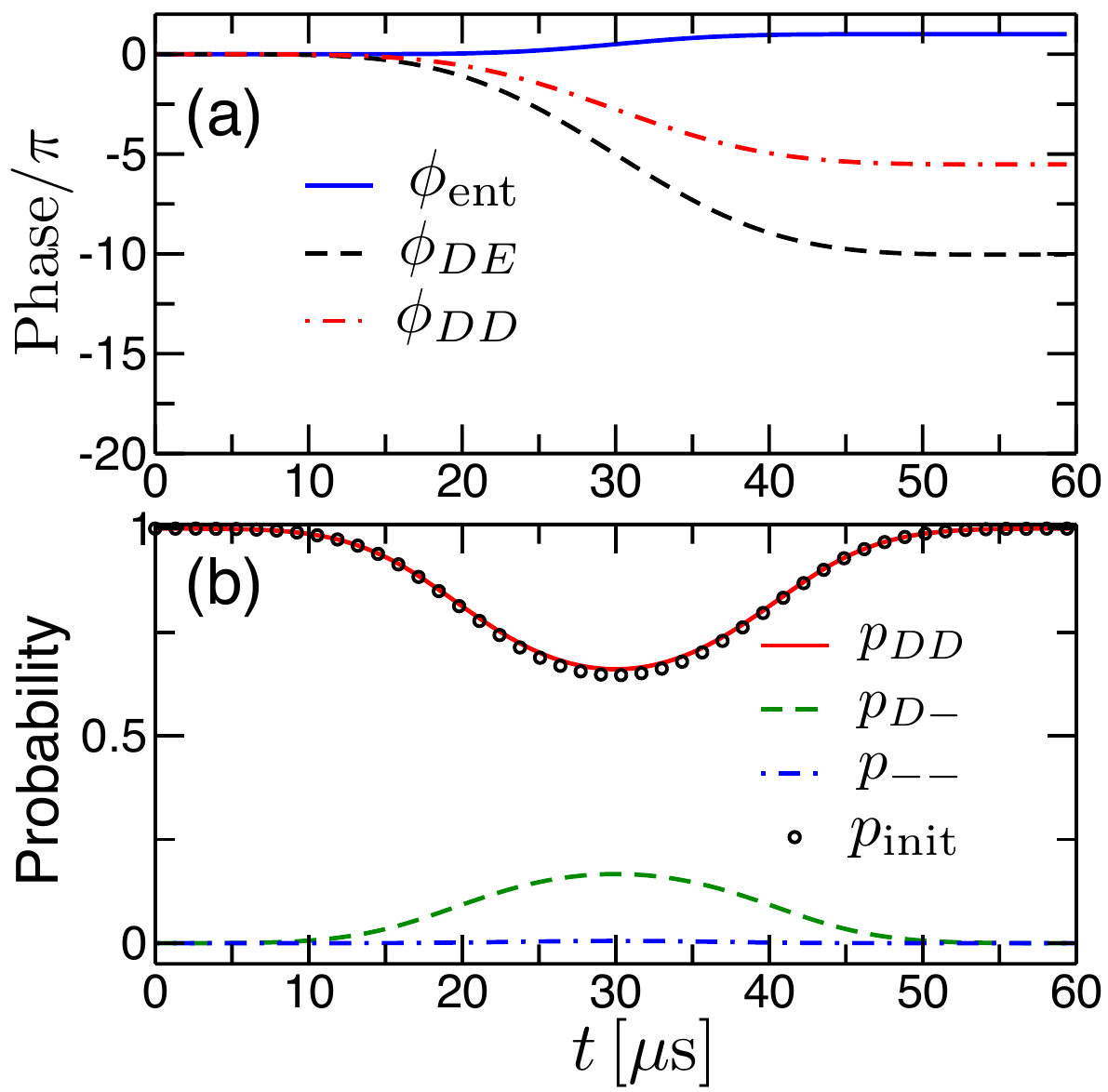}
\caption{(a) Phase evolution of different ion pair states. The parameters are chosen such that after the application of the laser pulse the entangling phase is $\phi_{\rm{ent}}=\pi$. (b) Excitation probability of certain ion pair states during the laser pulse when starting from the state $|DD\rangle\otimes|0\rangle$. By tracing out the CM phonon states (the maximal CM phonon number used in the simulation is 5), we obtain the probability $p_{DD}$ of the state $|\rm{DD}\rangle$, $p_{D-}$ of the state $|D-\rangle$ and $p_{DD}$ of the state $|--\rangle$. The probability of remaining in the initial state is $p_{\rm{init}}$. The parameters used for calculating the data are: The interaction energy $C_3(-)/R_0^3=2\pi\times 2.5$ MHz, $\omega_z=2\pi\times 1$ MHz and Lamb-Dicke parameter $\eta=0.5$. The remaining parameters are given in the caption of Fig.~\ref{fig:gate_laser}.}
\label{fig:phasepopulation}
\end{figure}

When neglecting the phonon dynamics the adiabatic unitary evolution of the qubit states under the Hamiltonian (\ref{eq:gatehamiltonian}) can be calculated analytically~ \cite{jaksch_fast_2000}. Using the two-ion state basis $\{|EE\rangle,\,|DE\rangle,\,|ED\rangle,\, |DD\rangle\}$, one finds that it implements the following phase rotation:
\begin{equation}
U_{\rm{gate}}=\left(\begin{array}{cccc}1 & 0 &0 &0 \\ 0 & e^{i\phi_{DE}}& 0 & 0\\ 0& 0&e^{i\phi_{DE}}& 0\\ 0 & 0& 0 & e^{i(\phi_{\rm{ent}}+2\phi_{DE})}\end{array}\right)
\end{equation}
Here the entangling phase is given by $\phi_{\rm{ent}}=\phi_{DD}-2\phi_{DE}$ where $\phi_{DD}=\int_{0}^{\tau}E_{\rm{DD}}dt$ and $\phi_{DE}=\int_{0}^{\tau}E_{\rm{DE}}dt$ are the accumulated phase of the $|DD\rangle$ and $|DE\rangle$ state, respectively~\cite{jaksch_fast_2000}. The adiabatic energies of the instantaneous eigenstates are
\begin{eqnarray}
E_{\rm{DD}}=\frac{1}{2}\left[\delta_0-\sqrt{\delta_0^2+2\Omega_-^2}\right],\\
E_{\rm{DE}}=\frac{1}{2}\left[E_--\sqrt{E_-^2+\Omega_-^2}\right],
\end{eqnarray}
with $\delta_0=E_--\Omega_-^2/(4E_-+2C_3(-)/R_0^3)$. The controlled phase gate is realized after removing the trivial phase $\phi_{DE}$ and $\phi_{ED}$ (via single qubit operation). Ideally the gate realizes an entangling phase $\phi_{\rm{ent}}=\pi$ which can be achieved by optimizing the laser parameter set $\{\Omega_0,\,\Delta_0,\,\tau\}$. One example of such optimal phase evolution is illustrated in Fig.~\ref{fig:phasepopulation}a.

Let us now take into account the phonon dynamics according to Hamiltonian Eq.~(\ref{eq:gatehamiltonian}). To get a qualitative idea about the effect of the phonons we investigate an idealized situation where we calculate the time-evolution of the initial state $|DD\rangle\otimes |0\rangle$ in which both ions are in the electronic state $|D\rangle$ and the CM phonon mode is not populated. In Fig.~\ref{fig:phasepopulation}b we display the time evolution of the populations of the electronic ion pair-states during the application of the gate laser pulse. We find that the population $p_{\rm{init}}$ of the initial state $|DD\rangle\otimes |0\rangle$ slightly deviates from the probability to remain in the state $|DD\rangle$, which is obtained by tracing out the CM phonons. The same behavior is found for other pair states which indicates that there is a slight population of the phonon states during the laser pulse. The magnitude of this phonon excitation is controlled by the trap frequency and can be reduced to an arbitrary degree if the confinement strength is increased.

Let us briefly discuss further gate errors caused by the spontaneous decay of ions from the Rydberg state. For $n=n'=65$, the lifetime of the MW dressed $|-\rangle$ state is $\tau_0\approx 132\,\mu$s. The corresponding loss from the pair states $|D-\rangle$ and $|-D\rangle$ can be estimated by $P_{\rm{loss}}\approx(2/\tau_0)\times \int_0^{\tau}p_{D-}(t)dt\approx 0.052$ with $p_{D-}(t)$ being the excitation probability in the state $|D-\rangle$. This loss can be further reduced by increasing the gate speed and considering even higher Rydberg states with longer lifetimes.

\section{Conclusions and Outlook}
\label{conclusion}
In conclusion, we have studied the implementation of a two-qubit phase gate in trapped ions which relies on the dipolar interaction between ionic Rydberg states. We have discussed in detail a number of technical difficulties which highlight central differences with respect to the implementation of similar gates among neutral atoms, and showed that they can be in principle overcome by utilizing MW dressed Rydberg states. Based on these dressed states we have briefly discussed the implementation of a controlled two-qubit phase gate and gave a first account on the effect of the electron-phonon coupling on the gate dynamics. In the future, it will be interesting to extend our analysis to larger ion crystals with thermal phonon states, in order to assess the usefulness of dipolar interactions for achieving a scalable ion trap quantum computer.

\acknowledgements
Discussions with all members of the R-ION consortium are kindly acknowledged. We thank C. Ates and S. Genway for careful reading of the manuscript. This work is funded through EPSRC and the ERA-NET CHIST-ERA (R-ION consortium). WL is supported through the Nottingham Research Fellowship by the University of Nottingham.

\end{document}